\documentclass[%
 reprint,
showpacs,
showkeys,
 amsmath,amssymb,
 aps,
]{revtex4-1}

\usepackage{graphicx}
\usepackage{dcolumn}
\usepackage{bm}

\begin{document}

\title{Estimation of mutual information for real-valued data with
  error bars and controlled bias}%

\author{Caroline M.\ Holmes}
\email{cholmes@princeton.edu}
\affiliation{Department of Physics, Princeton University, Princeton, NJ 08544, USA}
\author{Ilya Nemenman}%
 \email{ilya.nemenman@emory.edu}
\affiliation{Department of Physics, Department of Biology,\\ Initiative in Theory and Modeling of Living Systems\\ Emory University, Atlanta, GA 30322, USA}%

\date{\today}

\begin{abstract} 
  Estimation of mutual information between
  (multidimensional) real-valued variables is used in analysis of
  complex systems, biological systems, and recently also quantum
  systems. This estimation is a hard problem, and universally good
  estimators provably do not exist. Kraskov et al.~(PRE,
  2004) introduced a successful mutual information estimation approach
  based on the statistics of distances between neighboring data
  points, which empirically works for a wide class of underlying
  probability distributions. Here we improve this estimator by (i) expanding
  its range of applicability, and by providing  (ii) a  self-consistent way of
  verifying the absence of bias, (iii) a method for estimation of its
  variance, and (iv) a criterion for choosing the values of the free
  parameter of the estimator. We demonstrate the performance of our
  estimator on synthetic data sets, as well as on neurophysiological
  and systems biology
  data sets. \end{abstract}


\maketitle

\section{\label{sec:intro}Introduction}
Much of 20th century statistical physics was built by studying
dependences among physical variables expressed through their variances
and covariances. However, in recent decades, physicists have
started to explore systems (particularly those far from equilibrium),
where correlation functions, which are the most useful in the
context of small fluctuations and perturbative calculations, do not
tell the whole story about the underlying systems, which exhibit large,
nonlinear fluctuations. A related problem is that correlation
functions depend on the choice of a parameterization used to measure
observables, so that, for example, for large fluctuations, the
correlation between $x$ and $y$ can be very different from that
between $\log x$ and $\log y$, making it harder to interpret the data.

A common solution to these problems is to use the mutual information
between two variables instead of their correlation to quantify
dependence \cite{shannon,cover2012elements}. Mutual information
between variables $x$ and $y$ is distributed according to a joint
distribution $P(x,y)$ is defined as 
\begin{equation}
I_P[X,Y]=\int dx\, dy\, P(x,y) \log_2 \frac{P(x,y)}{P(x)P(y)},
\end{equation}
where the integral should be interpreted as a sum for discrete
variables, and as a multi-dimensional integral for multi-dimensional
real-valued variables. Mutual information quantifies {\em all}, and
not just linear dependences between the two variables: it is zero if
and only if the variables are completely statistically independent
\cite{cover2012elements}. Further, mutual information does not change
under invertible transformations (reparameterizations) of $x$ and $y$
\cite{cover2012elements}. These properties make mutual information the
quantity of choice for analysis of dependences between real-valued,
nonlinearly-related variables, especially in modern biophysics (see
Refs.~\cite{Fairhall:2012hk,Levchenko:2014dy,Tkacik:2016ch} for just a
few examples).

An important complication that prevents an even wider adoption of
information-based analyses is that mutual information and related
quantities are notoriously difficult to estimate from empirical data.
Mutual information involves averages of logarithms of $P$, the
underlying probability distribution. Since, for small $P$,
$-\log_2 P\to\infty$, the ranges of $x,y$ where $P$ is small and hence
cannot be sampled and estimated reliably from data contribute
disproportionately to the value of information. In other words, unlike
correlation functions, information depends nonlinearly on $P$, so that
these sampling errors result in a strong sample size dependent and
$P$-dependent bias in information estimates. In fact, even for
discrete data, there can be no universally unbiased estimators of
information until the number of samples, $N$, is much larger than the
cardinality of the underlying distribution, $K$
\cite{Paninski:2003vz}. This means that, for continuous
variables, universally unbiased information estimators do not exist at all. 
These simple observations have resulted in a lively field of developing
entropy / information estimators for discrete variables, which work
under a variety of different
assumptions (see
\cite{Panzeri:1996kv,strong1998entropy,Nemenman:2002tm,Paninski:2003vz,Panzeri:2007du,Zhang:2012ba,Berry:2013tn,archer2014bayesian}). Such estimators often use one of the following ideas. First, for $N\gg1$,
when most possible outcomes have been observed in the sampled data,
one may hope that the bias of an estimator can be written as a power
series in $1/N$, and then the first few terms of the series can be
calculated analytically, or estimated directly from data by varying
the size of the data set. Second, coincidences start happening in data
at much smaller $N$ than it takes to sample every possible outcome
\cite{Ma-1981}. One can then use
the statistics of such frequently occurring outcomes to extrapolate
and learn properties of the large low-probability tail of the
distribution $P$, estimating contributions of the tail to the
information. Third, one can estimate the bias of an estimator by
applying it to a shuffled data set, where the mutual information is
zero by construction. Some of these ideas can be applied to continuous
variables as well, by soft or hard discretization of the data.

However, for many experiments dealing with continuous variables, such
as when studying motor control, some of these bias correction
approaches are not easily applicable
\cite{tang2014millisecond,srivastava2017motor}. First, the observed
variables may be very large dimensional, which makes good sampling
nearly impossible. Second, when focusing on mutual information between
just two variables that are projections of very large dimensional variables, shuffling may not work as a way to check bias. Indeed, for
any finite $N$,  shuffling is not guaranteed to remove statistical
dependences among {\em all} data dimensions simultaneously, and randomizing along one set of projections may leave residual mutual information due to statistical dependences along the others.  Thus developing information estimators that use
continuity of real-valued data to help with undersampling, estimate
information without resampling, and work for large-dimensional data is
crucial. One of the most successful such estimators was proposed by
Kraskov, St{\"o}gbauer, and Grassberger \cite{kraskov2004estimating},
which we will refer to as KSG. It uses distances to the $k$-th
nearest neighbors of points in the data set to detect structures in
the underlying probability distribution. If some points cluster, then
the $x$ coordinate of a point can be used to predict its $y$
coordinate, resulting in a nonzero mutual information. This can be
detected by the statistics of the $k$-th nearest neighbor
distances. Further, by varying $k$, one can vary the spatial scale on
which structures are detected.

While successful, KSG cannot be a universally good for all underlying
probability distributions. In fact, even the original
Ref.~\cite{kraskov2004estimating} pointed out that there are
probability distributions for which the estimator does not converge
to the right answer even at very large $N$. However, we are not aware
of any published methods for self-consistently detecting if the
estimator is unbiased on specific datasets.  Our goal here is to make
KSG more broadly useful by endowing it with the abilities (i) to estimate
its own error bars, (ii) to detect existence of a sample-size
dependent bias, and (iii) to automatically choose the hyperparameter $k$
most appropriate for the current data. Further, (iv) we directly
expand the range of probability distributions, for which the estimator
remains unbiased, by using the reparameterization invariance property
of the mutual information.

Some of the methods presented in this paper were first tried in
Ref.~\cite{srivastava2017motor}, but here we test them more
thoroughly, introduce additional changes, and formalize the approach.
We start this paper with a brief review of the KSG estimator. We then
progressively introduce our modifications of the method. Finally we
give examples of performance of the modified method on simulated and
real-life data sets.

\subsection{\label{sec:estim}The KSG estimator}
Mutual information can be written down as the difference of marginal
and joint Shannon entropies \cite{cover2012elements}:
\begin{equation}
I(X,Y) = H(X) + H(Y) - H(X,Y).
\label{eq:MI}
\end{equation}
KSG uses the Kozachenko-Leonenko (KL) $k$th nearest neighbor entropy
estimator \cite{kozachenko1987sample} for each one of the differential
entropy terms:
\begin{equation}
\hat{H}_{\rm KL}(X) = -\psi(k) + \psi(N) + \log(c_d) + \frac{d}{N} \sum_{i=1}^N \log \epsilon^{(k)}(i).
\label{eq:KL}
\end{equation}
Here $\psi$ is the digamma function, $d$ is the dimensionality of $x$,
$N$ is the total number of samples, $c_d$ is the volume of a unit ball  with $d$ dimensions, and
$\epsilon^{(k)}(i)$ is twice the distance between the $i$'th data
point and its $k$'th neighbor. The intuition is that, if the distances
$\epsilon^{(k)}(i)$  are small, then the underlying probability
distribution is concentrated, and the corresponding differential
entropy is also small. Notice that the metric for calculating
distances has to be defined {\em a priori} to apply this estimator,
and the metrics can be very different in the $x$ and the $y$ spaces.

One could plug in Eq.~(\ref{eq:KL}) for each one of the three
differential entropies in Eq.~(\ref{eq:MI}), but then the biases in
the estimates of the marginal and the joint entropies likely will not
cancel -- if the ball with the radius $\epsilon^{(k)}(i)$
includes the $k$th nearest neighbor of the $i$th data point in the
$d(x)+d(y)$ dimensional space, then the ball of the same radius will
include a lot more data points in just $d(x)$ or $d(y)$ dimensions.
Reference~\cite{kraskov2004estimating} argued that keeping the ball
size rather than $k$ constant for the marginal and the joint entropy
would result in the decrease of the total mutual information bias.  To
implement this, KSG uses the $\max (\Delta x, \Delta y)$ metric to
define the distance between two points that are $(\Delta x,\Delta y)$
away from each other. It then defines the smallest rectangle in the
$(x,y)$ space centered at a point $i$ that contains $k$ of its
neighboring points. One then denotes by $\epsilon^{(k)}_x(i)$ and
$\epsilon^{(k)}_y(i)$ the $x$ and $y$ extents of this rectangle, and
by $n^{(k)}_x(i)$ and $n^{(k)}_y(i)$ the number of points such that
$||x_j-x_i|| \le \epsilon_x(i)/2$ or
$||y_j-y_i|| \le \epsilon_y(i)/2$, respectively. Then the mutual
information is estimated as
\cite{kraskov2004estimating,stogbauer2004least}
\begin{equation}
\hat{I}_{\rm KSG}^{(k)}(X,Y) = \psi(k) -1/k - \langle\psi(n^{(k)}_x) + \psi(n^{(k)}_y)\rangle + \psi(N),
\label{eq:KSG}
\end{equation}
where averaging is over the samples.  Note that, if
$\langle \psi( n^{(k)}_x)\rangle$ and $\langle \psi(n^{(k)}_y)\rangle$
increase, the mutual information estimate drops. This can be
understood intuitively as follows. First, recall that
$\psi(n)\to \log n$ for large values of the argument, and thus grows
with $n$. Since $\psi(n)$ is convex up,
$\langle \psi( n^{(k)}_x)\rangle$ is large when $n^{(k)}_x(i)$ are
narrowly distributed (and the same for $y$, respectively). But if
values of $n^{(k)}_x(i)$ (or $n^{(k)}_y(i)$) are nearly the same for
all $i$s, then the underlying probability distribution has no
structural features in the $x$ (or $y$) direction, and the mutual
information must be low, which is exactly what Eq.~(\ref{eq:KSG})
suggests.

Empirically, KSG is one of the best performing mutual information
estimators for continuous data. It has been used widely, with over 1700 citations to the original article according to Google Scholar as of the writing of this article. And yet some
basic questions remain unanswered. Foremost is that $k$ is a
free parameter, which needs to be chosen before applying the estimator
to data. Varying $k$ allows one to explore features in the probability
distribution across different spatial scales, resulting in the usual
bias-variance tradeoff. For example, $k = 1$ will pick up even very
fine features, but at the same time $n_x^{(k)}(i)$ and $n_y^{(k)}(i)$
will be small, resulting in large fluctuations.  On the other hand,
large $k$ may miss fine-scale features and hence underestimate the
information, but statistical fluctuations will be smaller. One can
expect that the optimal value of $k$ depends on the structure of the
spatial features in the data, which may be nontrivial and may exist on
multiple spatial scales. In addition, the optimal $k$ should also
depend on $N$, since fine features can only be observed at high sampling density.  Thus choosing the best $k$ is not a simple task. The
original KSG analysis focused largely on $N\to\infty$ and on
probability distributions with large, uniform spatial features, for which  $k\sim N$ was often useful (though $k = 2\dots4$, which is small but not 1, was also recommended).  In contrast, real life problems
often have $N\sim 10^2\dots10^4$ and many heterogeneous spatial
features, so that only $k\sim 1$ may have a chance of working. In this
article, in addition to other modifications, we propose a way of
estimating an optimal value of $k$ for KSG. Crucially, in order to do
so, we first solve two other problems: estimating the standard error
of the estimator and its bias directly from data.

\section{Results}
\subsection{Estimating the variance of KSG}

We first focus on estimating the standard deviation of KSG. For this,
we start with bivariate normally distributed data as a test case
since, for such data, the choice of $k$ has only a small effect on
$\hat{I}_{\rm KSG}^{(k)}$ \cite{kraskov2004estimating}. Additionally,
for a bivariate Gaussian, the true value of mutual information is
related to the correlation coefficient $\rho$ as
$I_{\rm true}=-\frac{1}{2}\log_2(1-\rho^2)$, which allows for an easy
determination of the actual error of the estimator. Specifically, for
the rest of this section, we will frequently use $\rho=0.6$ as an example, where
$I_{\rm true} \approx 0.32$ bits.

For a single data set taken at random from this bivariate Gaussian,
KSG will produce an estimate, e.~g., $0.2802$ bits for
$N=1000$. However, since we do not know the standard deviation of the
estimator (its ``error bars''), we do not know how many of these
digits are significant, and whether the estimate is biased.
Calculating the error bars is not simple since standard methods, such
as bootstrapping, only work for quantities that are {\em linear} in
the underlying probability distribution \cite{Efron:1993tv}, while
information is not. This is easy to understand intuitively: resampling
data with replacements -- a key step in bootstrap -- creates duplicate
data points. These will be interpreted by KSG as fine-scale,
high-information features, leading to overestimation of the mutual
information in the bootstrapped samples.

To illustrate the inadequacy of bootstrap for this problem, we
generate 20 independent sets of data of size $N=200$ from a bivariate
Gaussian with $\rho=0.6$. We then estimate $\hat{I}_{\rm KSG}^{(1)}$
for each set, and finally calculate the mean and the standard
deviation of these 20 KSG estimates. The result is
$\hat{I}_{\rm KSG}^{(1)} = 0.32 \pm 0.12$ bits, which matches well
with the analytical value of $\approx 0.32$ bits. On the other hand, if
we take the single data set of $N=200$ and then bootstrap it and
calculate the mean and the standard deviation of the KSG estimates of
the bootstrapped data, we get
$\hat{I}_{\rm KSG}^{(1)} = 1.32 \pm 0.21$ bits. The mean is wrong by a
factor of about 4, and even the standard deviation is twice as large
as it should be (and the scale of both errors certainly depends on $N$
and the underlying distribution). We emphasize this again: {\em
  bootstrapping, at least in its simple form, should not be used in
  estimation of mutual information or its error bars!}

\begin{figure}[!t]
\includegraphics[width = 3.4in]{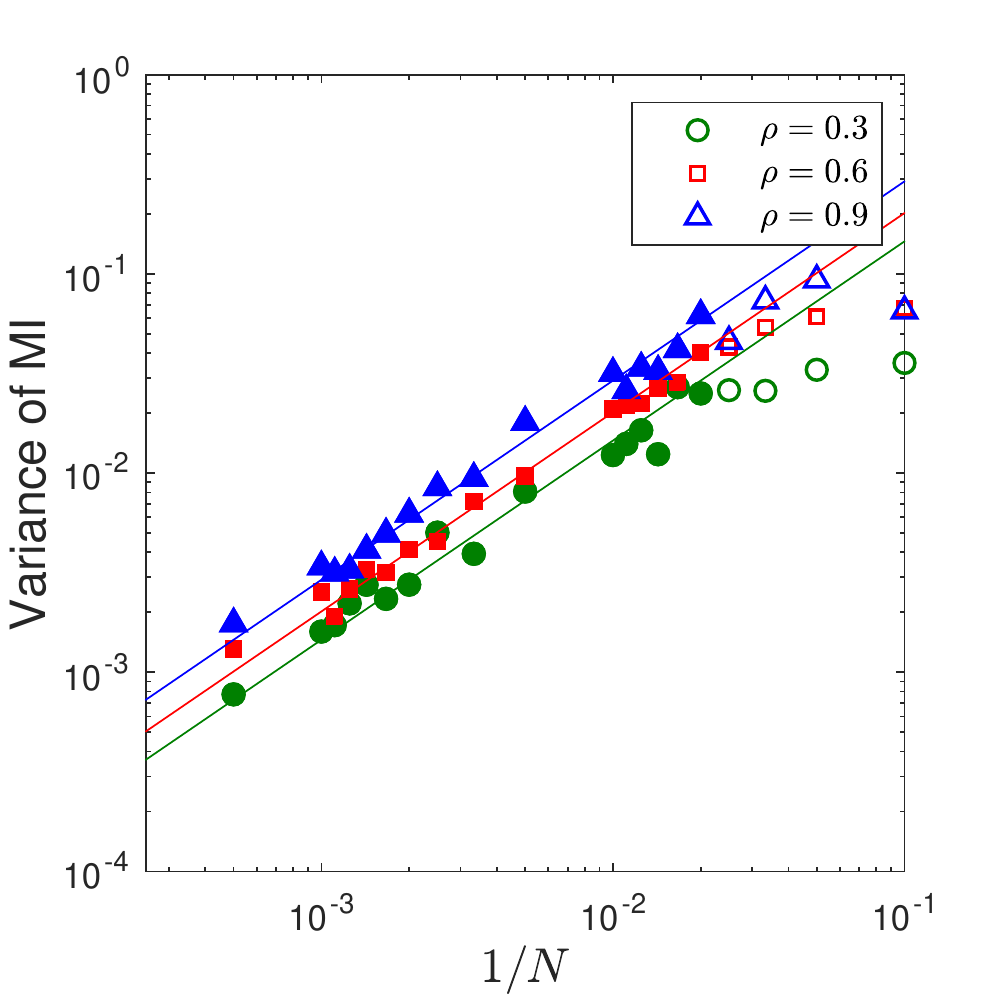}
\centering
\caption{{\bf Dependence of the variance of KSG on the sample set
    size.} For bivariate Gaussians with three different correlation
  coefficients $\rho= 0.3, 0.6, 0.9$, we generate $100$ independent
  sample data sets of different sizes $N$. For each $N$, we calculate
  $\sigma^2_{\rm KSG}(N)$ as the empirical variance of all
  $\hat{I}^{(1)}_{\rm KSG}$ with this $N$. The variance is plotted
  vs.\ $1/N$. The shown linear fit illustrates that the variance,
  indeed, scales as $1/N$ for $N\gg 1$. Empty symbols were not used to fit the linear relation. \label{fig:VarScaling}}
\end{figure}

Instead of using bootstrapping for estimating the error of KSG, we
propose to use the fact that variance of essentially any function
that, like Eq.~(\ref{eq:KSG}), is an average of $N$ random i.~i.~d.\
contributions scales as $1/N$ for sufficiently large $N$. Indeed, as
seen in Fig.~\ref{fig:VarScaling}, this scaling holds, for example,
for bivariate Gaussians with different correlation coefficients for,
at least, $N > 50$.

Thus we write for the variance of KSG 
\begin{equation}
\sigma_{\rm KSG}^2(N) = \frac{B}{N},
\label{varScaling}
\end{equation}
where the value of $B$ will depend on the particular distribution. To
estimate $B$ for specific data, we subsample ({\em not} re-sample!)
the data. Specifically, for a small integer $n$, we partition the data
set of size $N$ at random into $n$ non-overlapping subsets of as close to
equal sizes as possible. We calculate $\hat{I}^{(k)}_{\rm KSG}$ for each such
subset. Then the sample variance of these $n$ values of
$\hat{I}^{(k)}_{\rm KSG}$ is our estimate of
$\sigma_{\rm KSG}^2(N/n)$. Once we know $\sigma_{\rm KSG}^2(N/n)$ for
many values of $n$, we fit the model, Eq.~(\ref{varScaling}), to these
values and estimate $B$ empirically. Finally, knowing $B$, we
calculate $\sigma_{\rm KSG}^2(N)$ from Eq.~(\ref{varScaling})
directly.  Combining these steps, we get expressions for the estimate of the
variance of the estimator, as well as the standard error of the
variance itself, which can be found in the Appendix, Eqs.~(\ref{eq:var_est}) and (\ref{eq:var_var}), respectively.

\begin{figure}[!t]
\includegraphics[width = 3.4in]{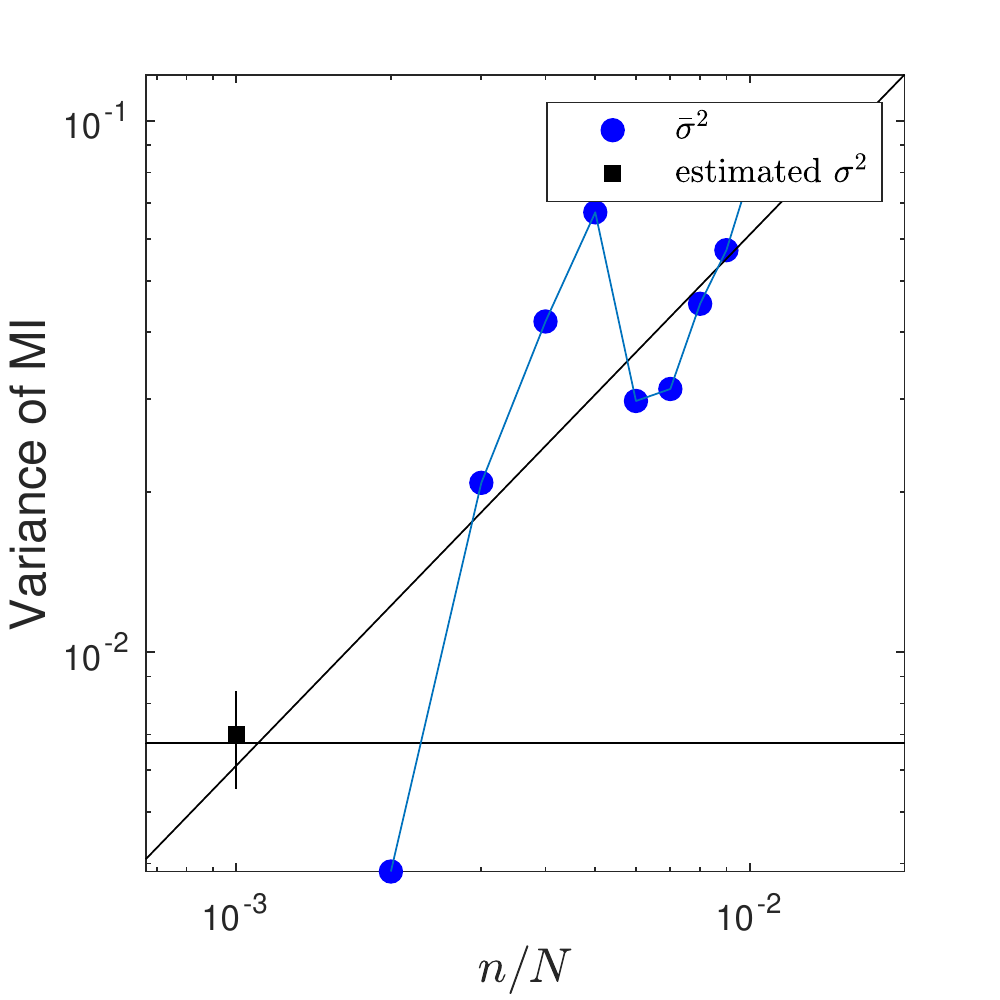}
\centering
\caption{{\bf Calculating the variance of KSG.} For a bivariate
  Gaussian with $\rho = 0.6$, we sample $N = 1000$ data points from
  the distribution. We calculate the variance of KSG with $k=1$ for $N/n$ data points by partitioning the data into $n$
  nonoverlapping subsets and estimating the mutual information for
  each subset, as described in the main text (blue dots). An
  unweighted linear fit with the slope of 1 is shown as a guide to
  eye, illustrating extrapolation of the variance of the estimator to
  the full data set size. An estimate of the variance of the estimator, with its own
  expected error, is performed using the analysis in the Appendix and is
  denoted by a black square with an error bar. For comparison, the
  horizontal line denotes the variance of the estimator calculated
  from applying it to 100,000 independent samples of size
  $N=1000$ from the Gaussian, illustrating a near perfect agreement.
  \label{fig:VarExtrap}}
\end{figure}

We finish the Section with a few observations. First, one might be
tempted to generate many different non\-overlapping partitions of the
data at the same $n$, hoping to average over the partitions and hence
decrease the variability observed in Fig.~\ref{fig:VarExtrap}. This should
be avoided since such different permutations of data would not produce
independent samples of the variance. For the same reason, one should
avoid any overlaps among partitions, so that the number of samples in
each partition is $N/n$ with an integer $n$.  Finally, the $1/N$
scaling of the variance only works for large $N$.  Thus it may not
hold for $n\gg 1$, limiting the maximum value of $n$ in realistic
applications. For all plots shown here, we use $n = 1\dots10$, which we generally find to be sufficient.

\subsection{Detecting the estimation bias and choosing $k$}
Most common mutual information estimators, including KSG, are
asymptotically unbiased for sufficiently regular probability
distributions at $N\to\infty$. At the same time, all are typically
biased at finite $N$, as discussed in the Introduction. As a
result, the bias is {\em sample size dependent}. Thus while it may be
hard to calculate the bias analytically for specific data and
estimators, one may be able to estimate it empirically by varying the
size of the data set
\cite{strong1998entropy,Nemenman:2008ft,tang2014millisecond,srivastava2017motor}:
if the estimated mutual information drifts with changing $N$, there
are reasons to be concerned about the bias. Here we will use this
strategy to ascertain the existence of a sample size dependent bias
for KSG.

We note that, unlike Ref.~\cite{strong1998entropy}, we are not
interested in estimating the bias at finite $N$ and then subtracting
it out (equivalently, extrapolating $\hat{I}^{(k)}_{\rm KSG}$ to
$N\to\infty$). This is possible only when the form of the bias as a
function of $N$ is known, leaving only a small number of coefficients
to be characterized from data themselves, such as for the classical
$\sim 1/N$ Miller-Madow correction to the maximum likelihood
information estimator \cite{miller}. For KSG, the
asymptotic scaling of the bias is unknown, making this approach
currently infeasible. Further, any estimator would exhibit statistical
fluctuations when applied to real data. Unless the standard deviation
of the estimator is known, one cannot say whether the observed sample
size dependent drift is due the bias or to the fluctuation: only if
the systematic drift over a reasonable range of $N$ is much larger
than the standard deviation, would one consider this an evidence of
the bias. Thus detecting the bias of KSG (or any other estimator) by
varying $N$ is impossible without a careful consideration of how
$\sigma^2_{\rm KSG}$ behaves.

 The question of detecting the bias is
intimately related to choosing $k$, the number of nearest neighbors
considered by the estimator: we expect the bias to be $k$-dependent.
Specifically, for large $k$, fine-scale features in the underlying
probability distribution will be missed by KSG, and the mutual
information will typically be underestimated. At the same time,
because $n_x^{(k)}$ and $n_y^{(k)}$ grow with $k$, we expect the
standard deviation of the estimator to be smaller at larger $k$. In
contrast, for smaller $k$, statistical fluctuations will be much
larger, while two different effects will affect the bias. First, the
downwards information bias is expected to be smaller at small $k$
since finer scale features will be explored. Second, larger
fluctuations in $n_x^{(k)}$ and $n_y^{(k)}$ will lead to a larger
$N$-dependent upwards bias in $-\langle\psi(n_x^{(k)})\rangle$ and
$-\langle\psi(n_y^{(k)})\rangle$ in Eq.~(\ref{eq:KSG}). Overall, the
bias at small $k$ may be of an arbitrary sign. In any case, one can
explore the drift as a function of $N$ for different values of $k$ and
choose to work with the value (if one exists), for which (a) there is
no sample-size dependent drift compared to the estimator standard
deviation, and (b) the standard deviation is the smallest. We also note that the actual estimated value of the mutual information can be strongly $k$-dependent; we will discuss this further below, but we note here briefly that it is important for the estimated value of the information to be stable across a range of $k$'s. 

\begin{figure}[!t]
\includegraphics[width = 3.4in]{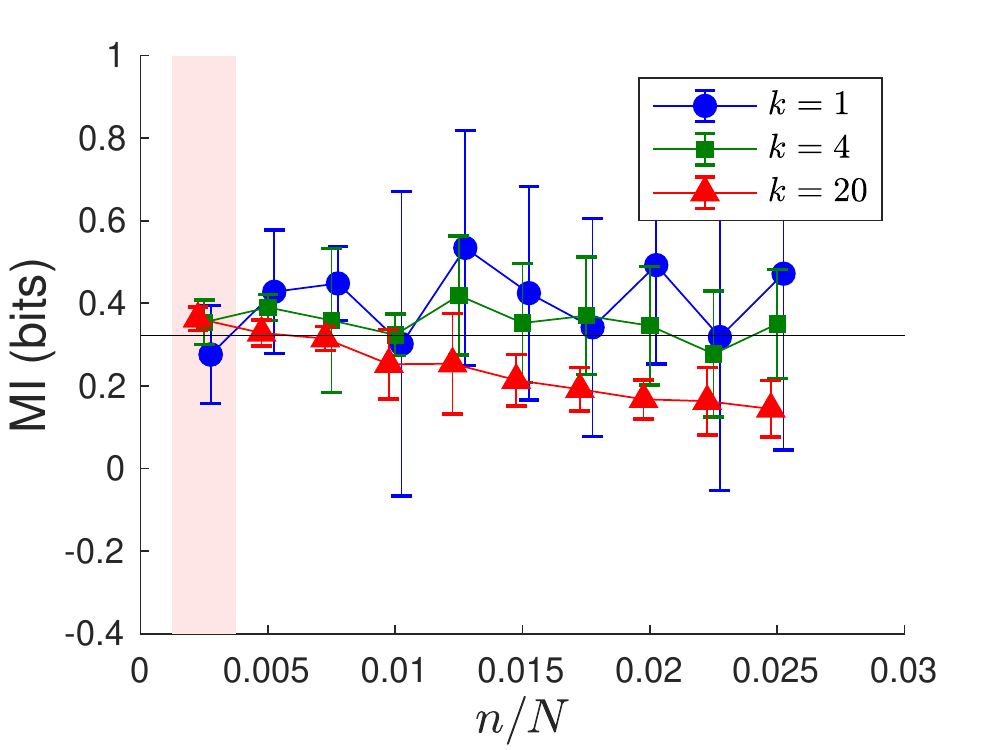}
\centering
\caption{{\bf Bias of KSG as a function of $N$ and $k$.}  Starting
  with $N=400$ samples from a bivariate normal distribution with
  $\rho=0.6$, we partition the data into $n$ nonverlapping subsamples
  (without replacements), each with $N/n$ data points. We estimate
  $\hat{I}^{(k)}_{\rm KSG}$ for each subsample using
  Eq.~(\ref{eq:KSG}). Means and standard deviations of the estimates
  for each set of $n$ partitions are shown for three different values
  of $k$. The leftmost point (on pink packground) for each line has error bars representing our estimate, following the methods we discussed in the previous section. The true mutual information of $0.322$ bits is shown as a
  black horizontal line. For the data set sizes explored here,
  $N/n=40\dots400$, $k = 20$ clearly leads to a statistically
  significant negative bias, while $k = 1$ gives an unnecessarily high
  variance, sometimes dipping into mathematically impossible negative
  values. $k=4$ shows a low-bias, low-variance behavior for these
  $N/n$. Note that symbols for different values of $k$ are slightly shifted relative to each other for visibility, but are actually evaluated at the same $n/N$ for all $k$.
  \label{fig:biasVsk}}
\end{figure}

We illustrate this analysis in
Fig.~\ref{fig:biasVsk} for the bi-variate normal distribution. Here we
work with smaller data sets than in the previous figures to better
explore the effects of $k$. Of the three values of $k$ shown in
the Figure, $k=4$ shows the best combination of no sample size
dependent drift and low variance. Correspondingly, as this drift
analysis predicts, $\hat{I}^{(4)}_{\rm KSG}$ remains unbiased compared
to the true mutual information value over the entire range of data
explored. We also verified that the estimator is relatively stable to
the choice of $k$, so that other values near $k=4$ give similar
$\hat{I}^{(k)}_{\rm KSG}$, and the estimator remains unbiased (not
shown).

\begin{figure}[!t]
\includegraphics[width = 3.4in]{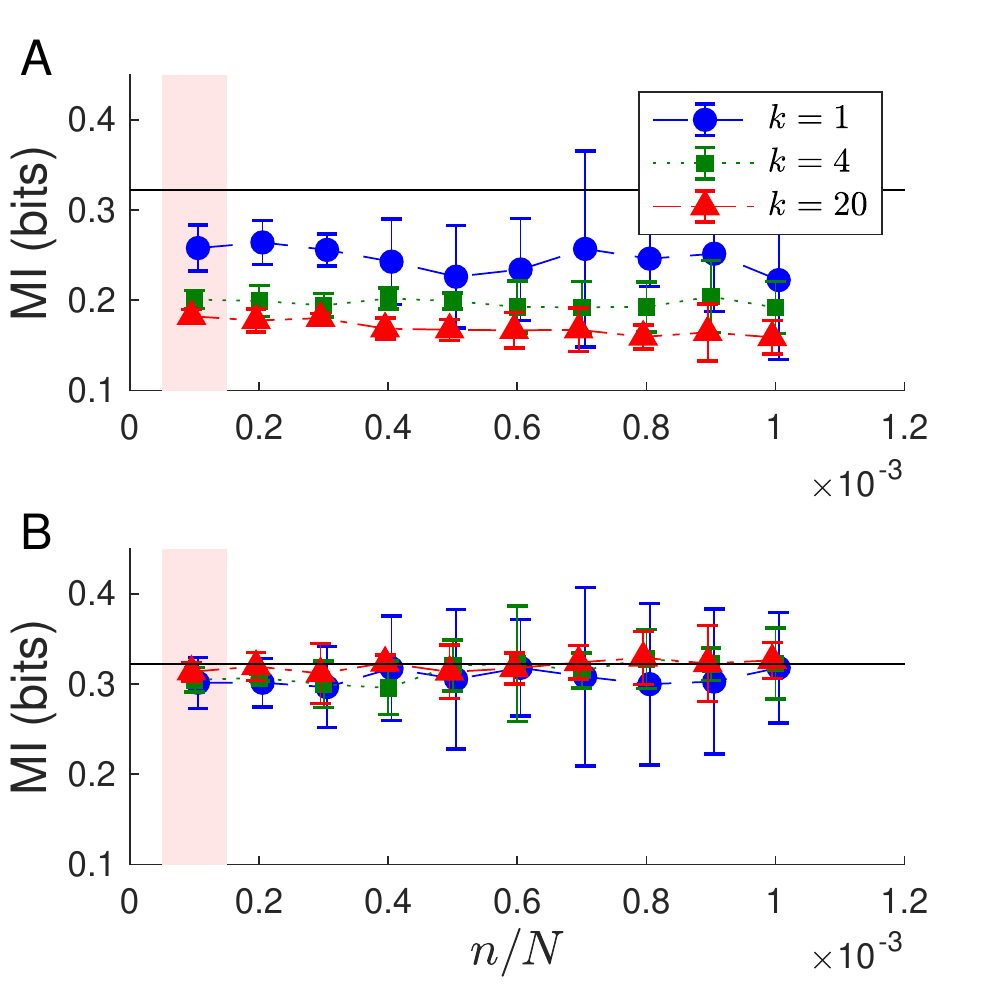}
\centering
\caption{{\bf Marginally normalizing the data decreases the KSG bias.}
  (A) For a bivariate log-normal distribution, $P(x,y) \sim \exp{(-((\ln 3x)^2 + (\ln 5y)^2 - 2 \rho \ln 3x \ln 5y )/(2(1-\rho^2)))}$, with $x$ and $y$ being standard normal,  $\rho=0.6$, and the true mutual information of  $0.322$ bits, we repeat the analysis from Fig.~\ref{fig:biasVsk} and
  plot the dependence of  $\hat{I}^{(k)}_{\rm KSG}(X,Y)$ on $k$ and $n/N$
  for $N = 10^4$. As always, the error bars on the leftmost points (full data set, pink background) are estimated as discussed above. The true value of information is
  shown as a black horizontal line. KSG does not give a consistent estimate of the information, and any estimate would be a function of $k$. No value of $k$ gives the correct mutual information. (B) After reparameterizing each marginal
  into a standard normal, we investigate the dependence of
  $\hat{I}^{(k)}_{\rm KSG}(X',Y')$ on $k$ and $n/N$. Here KSG does not
  show a sample sign dependent drift and is, therefore,
  largely unbiased for all tested values of $k$. Here we also have an estimate that is independent of the choice of $k$. \label{fig:Normalizing}}
\end{figure}

We note that Ref.~\cite{kraskov2004estimating} explored, in
particular, $k \propto N$, and $N\to\infty$. In contrast, our approach
often gives $k\sim 1$ for $N\sim 10^2\dots 10^4$. We expect that
$k\propto N^\eta$ for some distribution-dependent $\eta<1$ to be
asymptotically optimal since it would lead to both (i) exploring
progressively finer features and (ii) smaller relative fluctuations in
$n_x^{(k)}$ and $n_y^{(k)}$ as $N\to\infty$.  However, here we are
interested in applications to real experimental data sets. These are
usually far from the asymptotic regime, so that the available range of
$N$ is too small to meaningfully think about different scalings of
$k$.

\subsection{Decreasing the KSG bias}
Empirically, KSG exhibits large biases for distributions that have
very heavy tails, have structural features on multiple length scales,
or are severely skewed. All of this can be traced to the non-symmetric
distribution of data points in the $\epsilon$-balls. As an example, Fig.~\ref{fig:Normalizing} (A) shows
application of KSG for different values of $k$ to a bivariate
log-normal distribution. Even for a very large  $N = 10000$, KSG is
severely negatively biased for all $k$s. In specific realizations, we often see the bias {\em increasing} as $N$ grows, so that the KSG estimate turns negative, while mutual information must
always be positive. We note that small negative values
of information would not be a concern generally:  in order to estimate
information near zero bits with error bars, one needs to have it be
negative sometimes --- negative estimates that fall 
within error bars of zero are acceptable. Here, however, the estimates
can be consistently and significantly negative, indicating a serious
problem.

However, as we mentioned above, mutual information is invariant under
invertible marginal reparameterizations. Thus one can hope to increase
the range of distributions for which KSG is unbiased, by
reparameterizing the data to distributions that KSG is better equipped
to handle. Specifically, since KSG works extremely well for normal
variables \cite{kraskov2004estimating}, we suggest to transform each
marginal variable $x$ and $y$ into a standard normal variable. For
example, if we define $r_i=1\dots N$ as the rank of the corresponding
$x_i$, then its reparameterized version is
\begin{equation}
x'_i= \sqrt{2}\; {\rm Erf}^{-1} \left(2r_i - \left(N+1\right)\right),
\label{eq:reparam}
\end{equation}
where ${\rm Erf}^{-1}$ is the inverse of the error function. Indeed,
as illustrated in Fig.~\ref{fig:Normalizing} (B), this
transformation removes the bias for many cases. Note that we did not use the fact
that the distribution is bivariate log-normal during the
reparameterization: Eq.~(\ref{eq:reparam}) will transform {\em any} data
into marginally normal variables.

In some sense, the log-normal example is trivial, since marginal
reparameterizations transform it not just into {\em marginally}
normal, but into {\em jointly} normal distribution, which would not be
expected generically. However, since KSG depends largely on {\em
  marginal} neighborhoods, cf.~Eq.~(\ref{eq:KSG}), one would expect
that joint normality after reparameterization is not necessary, and
marginal normality alone is sufficient for the bias to be
decreased. Below we illustrate this on two real experimental datasets.  

\begin{figure}[!t]
\includegraphics[width = 3.4in]{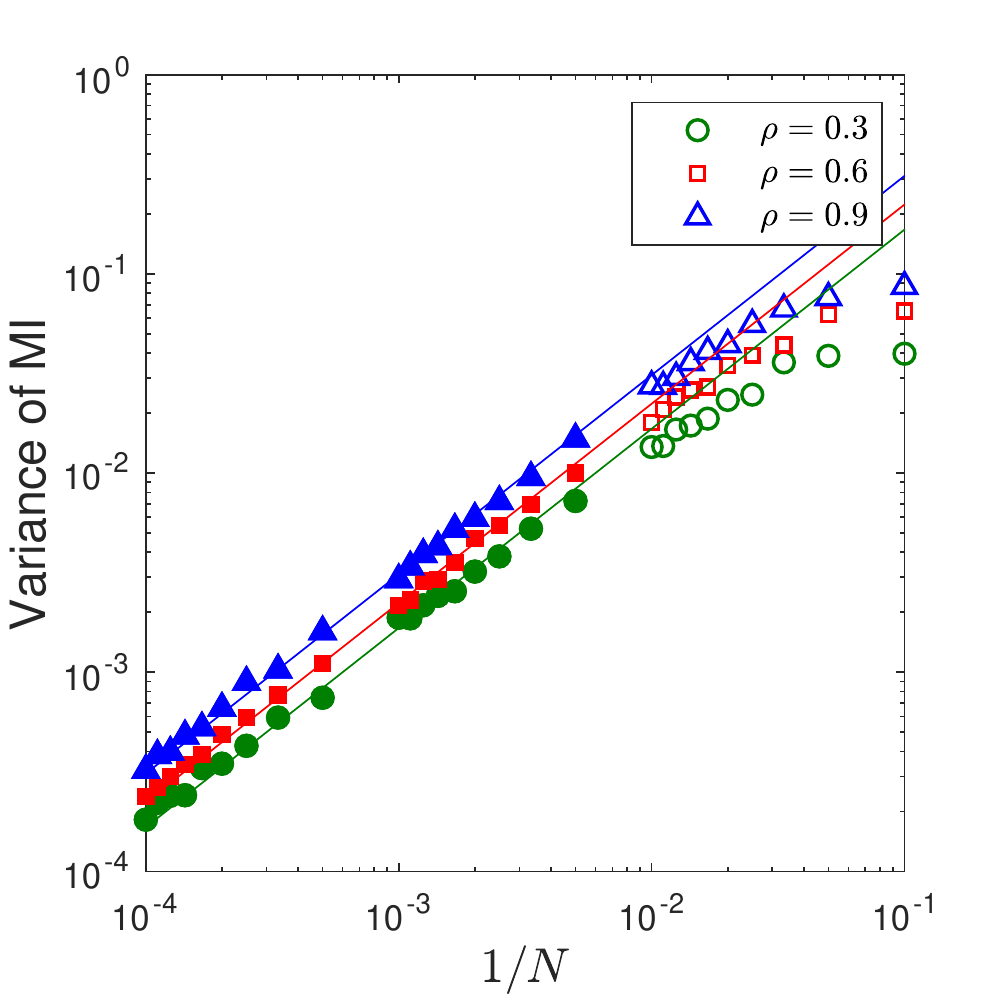}
\centering
\caption{{\bf Dependence of the variance of KSG on the sample set
    size for non-normal data}. We repeat the analysis of
  Fig.~\ref{fig:VarScaling}  for the reparameterized log-normal data of
  Fig.~\ref{fig:Normalizing}(B), as well as a few other
  log-correlation coefficients. Here, we have reparameterized from a skewed, heavy tailed distribution, which had biased information estimates. Nonetheless,
  the scaling $\sigma^2_{\rm KSG}\propto 1/N$ still holds, as
  illustrated by straight line fits, which have slopes of exactly 1. Empty symbols were not used to fit the straight lines.
\label{fig:varScalingNR}}
\end{figure}

However, before that, we need first to show that our procedure for estimating the variance
of the estimator can be used for reparameterized
data, where biases may exist, and where the original distribution is non-gaussian. For this, we repeat the analysis of Fig.~\ref{fig:VarScaling}
for reparameterized data: Figure \ref{fig:varScalingNR} shows
scaling of the KSG variance as a function of $N$ for the
reparameterized log-normal data,
cf.~Fig.~\ref{fig:Normalizing}(B). While the mutual information
estimate on the underlying distribution is severely biased, with our reparameterization we are able to not only return to a regime where we can make unbiased estimates, but also where we have the $1/N$ variance scaling. \\

\begin{figure}[!t]
\includegraphics[width = 3.4in]{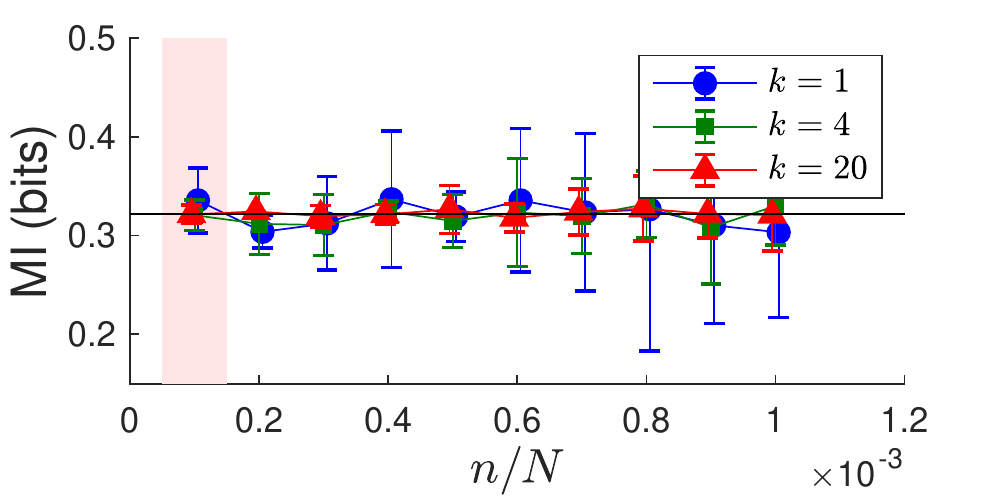}
\centering
\caption{{\bf KSG for multivariate data.}  While other choices could be explored, we chose to start with the same log-normally
  distributed data as in Fig.~\ref{fig:Normalizing}. We then rotate
  $x$ into three components,
  $(x_i^{(1)}, x_i^{(2)}, x_i^{(3)})= x_i\times (\cos \phi \cos
  \theta, \sin\phi \cos\theta, \sin \theta)$, where $\phi= \pi/6$ and
  $\theta = \pi/3$. We similarly make $y$ three dimensional with
  the same $\phi$ and $\theta$. Now KSG needs to find the information
  between two three-dimensional log-normal variables. For these data,
  KSG is biased (not shown). However, performing marginal reparametereizations for each of
  the six involved variable components independently, we recover the
  unbiased performance statistically indistinguishable from
  Fig.~\ref{fig:Normalizing}:  the KSG estimate does not show
  sample size dependent drift, is consistent for many $k$s, and
  matches the analytical information value (black horizontal line) for the full data set (pink background).
  \label{fig:multidim}}
\end{figure}
A similar reparameterization prescription works for estimating mutual information between
higher dimensional variables, although the problems of undersampling
are amplified in this case. We first transform each component of the
data into a standard normal variable using Eq.~(\ref{eq:reparam}). We
then estimate the estimator variance by performing a linear fit to
variances of partitions and then extrapolating to the full data set
size. Finally, we check for the $N$-dependent drift for various $k$,
and hence choose a good value of $k$, if one
exists. Figure~\ref{fig:multidim} shows application of the approach to
a 6-dimensional multivariate normal distribution (three dimensions
each for $x$ and $y$). As in the one-dimensional case, the
estimator does not work without reparameterization (not shown), but
it performs quite well for the marginally normalized data 
despite having to deal with more dimensions.

\section{\label{sec:guide}Practical Guide}

MatLab package for performing all of the analyses described above are
available from
\url{https://github.com/EmoryUniversityTheoreticalBiophysics/ContinuousMIEstimation}. In
this section, we describe functions in this package, list our specific
recommendations for using it to estimate mutual information for
continuous variables, and demonstrate how to do so using two
experimental data sets.

\subsection{Functions in the software package}

\verb!MIxnyn.m! We distribute the original KSG software (written in C
and MatLab) together with our modifications of it. Details for
compiling and installing the package are available in the {\tt README}
file. This function provides the MatLab interface to the C
implementation of KSG. It takes two vectors of samples $x_i$ and $y_i$
as input, where either or both can be multi-dimensional, assumes the
usual Euclidean metric on both the $X$ and the $Y$ space, and produces
a single estimate of the mutual information between the two variables.

\verb!findMI_KSG_subsampling.m! This function calculates
$\hat{I}^{(k)}_{\rm KSG}$ for the full data and its nonoverlapping
subsets. It takes two vectors of (potentially multi-dimensional)
samples $x_i$ and $y_i$ on the input, as well as a single value of $k$
and the vector of $n$, the number of subsets to divide the data
into. For each value in the vector $n$, it partitions the data into
this many nonoverlapping partitions at random, calculates
$\hat{I}^{(k)}_{\rm KSG}$ for each subset, and outputs results of all
of these calculations. It can additionally make a figure similar to
Fig.~\ref{fig:biasVsk} for a single value of $k$,
which allows the user to check for the sample-size dependent drift visually.

\verb!findMI_KSG_stddev.m! 
This function calculates the variance $\sigma^2_{\rm KSG}$ for the
full data set, as described above. For this, it takes the output of
\verb!findMI_KSG_subsampling.m!  (the mutual information values for
different subsamples of the data) as well as the data set size $N$ as
the input. It then calculates the sample variance of $n$ values of
$\hat{I}^{(k)}_{\rm KSG}(N/n)$ for all available $n$ and extrapolates
the variance to the full data set size of $N$. If requested, the
function can produce a figure similar to Fig.~\ref{fig:VarExtrap},
illustrating the procedure and allowing for a visual inspection of
whether the variance of subsamples is $\propto 1/N$, as expected.

\verb!findMI_KSG_bias_kN.m!  This is the
wrapper function that performs our analysis for different values of
$k$. It takes the $x_i$ and $y_i$ samples, the list of $k$s to try,
and the list of the number of data partitions $n$ as the input. It
calls the two previous functions sequentially and estimates
$\hat{I}^{(k)}_{\rm KSG}(N)$ with error bars for every value of
$k$. The function can additionally make a figure similar to
Fig.~\ref{fig:biasVsk} for all values of $k$ to help find the value
$k$ for which KSG has the smallest sample size dependent drift and
the smallest variance. The function outputs a list of mutual
information values with error bars, each corresponding to a specific
value of $k$.

\verb!reparamaterize_data.m! The function reparameterizes the data to
a standard normal distribution, which, if performed before other
estimation steps, should increase the range of applicability of
KSG. It takes a vector of samples $x_i$, which must be one
dimensional, as the input and returns the reparameterized data as the
output.
\begin{figure}[!t]
\includegraphics[width = 3.5in]{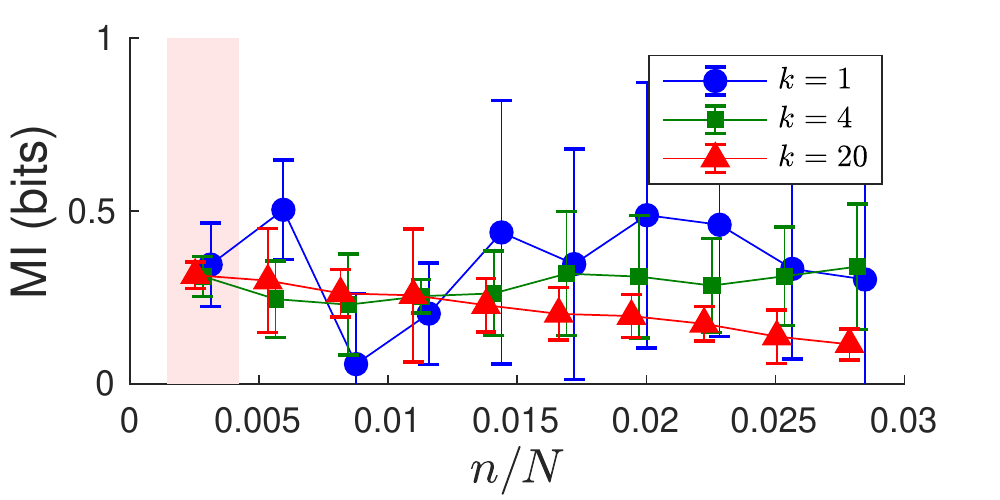}
\centering
\caption{{\bf Application of KSG to systems biology data.} Mutual
  between NF-$\kappa$B and p-ATF-2 activation in mouse fibroblasts 30
  min after activation with TNF at 1.3 ng/mL is shown. Data has been marginally
  reparameterized to a standard normal for this plot (without the
  reparameterization, estimates are  biased). For the full data set size (pink background), standard deviations are extrapolated as detailed above. $k=20$ shows
  downwards bias for, at least, large number of partitions. $k=1$ is
  unnecessarily noisy. $k=4$ exhibits a good balance of low drift (bias) and
  low variance. 
\label{fig:nfkb}}
\end{figure}

\subsection{Application notes}

\begin{enumerate}
\item Transform each of the components of both $X$ and $Y$ into the
  standard normal form using \verb!reparamaterize_data.m!. This should
  not have any negative effects on the estimation, and may turn out to
  be extremely advantageous. 
\item Do not use bootstrapping and related techniques to estimate
  variance of the estimator. 
\item For a few values of $k$, explore the dependence of the estimates
  on $k$ and the data set size using \verb!findMI_KSG_bias_kN.m! or
  other functions in the package. Look for a signature of the
  estimator drift for smaller data set sizes (many partitions), and
  similarly look for a signature of deviation from $\sim1/N$ scaling
  for the variance. These deviations and drift will set the maximum number of
  data partitions one can explore, and hence will limit the ability to
  verify whether the estimator is unbiased.
\item Choose the value of $k$ for which the estimator shows no
  statistically significant drift over the largest range of the data
  set size. If many such $k$s exist, choose the value for which the
  estimator error bars are the smallest over the range.  Note that the estimator should be stable in some range of $k$ around the optimal value, but one cannot expect the estimate to be fully independent of $k$.
\item Resist the temptation of subtracting the bias (extrapolating the
  estimator to $N\to\infty$), or declaring the estimator unbiased
  based only on a small range of $N$. Empirically, about a decade of
  stability in $N$ is needed for this determination. Recall that no
  estimator is universally unbiased, and so it might be impossible to
  estimate the information reliably from your data using KSG.
\item If no unbiased $k$ is found, try to reduce the dimensionality of
  your data by any available dimensionality reduction approach. Biases
  decrease rapidly when the dimensionality decreases. On the other
  hand, performing any manipulations with data cannot increase the
  information (by the Data Processing Inequality), and thus one may be
  able to estimate the lower bound on the true information reliably,
  with little bias, which may be sufficient for some applications.
\end{enumerate}

\subsection{Examples}
Our software package includes two experimental data sets, showing the
utility of the method and allowing one to practice estimation for 
realistic data.

The first data set comes from the systems biology literature and can
be found in \verb!NFkappaBData.mat!. These data were taken with
permission from Ref.~\cite{Cheong:2011jp}. The data describe the joint
activity of two transcription factors NF-$\kappa$B and p-ATF-2
measured in 335 individual wildtype mouse fibroblast cells 30 min
after exposure to the tumor necrosis factor (TNF) ligand at the
concentration of 1.3 ng/mL. The two transcription factors are
activated downstream of the same TNF receptor, and hence their
activity is correlated. The mutual information between these two sets
quantifies this relation. Figure \ref{fig:nfkb} shows application of
our method to these data. The figure can be generated by
\verb!NFkappaBDataExample.m!, which is included in the
distribution.

\begin{figure}[!t]
\includegraphics[width = 3.5in]{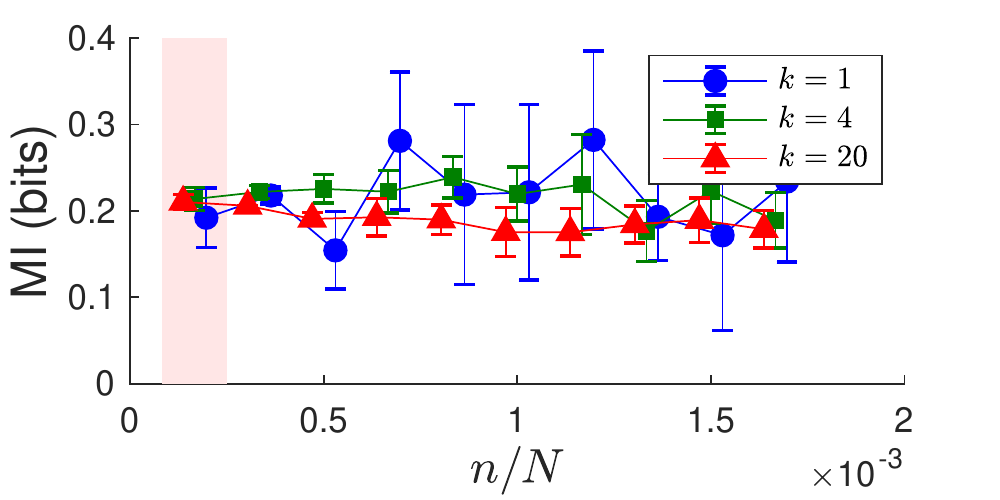}
\centering
\caption{{\bf Application of KSG to neurophysiological data.}  Mutual information between a pair of interspike intervals and the following pair of interspike intervals within a breathing cycle for anesthetized Bengalese finches is being estimated. Despite the high dimensionality and the non-Gaussian nature of the data, we are able to find a stable estimate for the information with $6000$ samples. The estimate is stable for many values of $k$, with similar error bars for $k>1$ ($k=1$ again gives unnecessarily large error bars). The unreparameterized case (not shown) performs markedly less well.
\label{fig:birds}}
\end{figure}

The second data set illustrates application of KSG to neurophysiology
data and can be found in \verb!BirdSpikingData.mat!.  The data have been
taken with permission from Ref.~\cite{srivastava2017motor}. They represent recordings of neural activity from anesthetized Bengalese finches, measured in the motor neurons that control breathing. Here we are analyzing the structure of the spike train itself. The recorded neurons fire only during a particular phase of the breathing cycle, and we are looking at the interspike intervals within such bursts. Specifically, we are estimating the mutual information between two subsequent interspike intervals as one variable, and the following two interspike intervals as the other. Importantly, this is high-dimensional (two dimensions for both $x$ and $y$) and non-Gaussian real data. Without reparameterization, questions would remain about the persistent bias of the estimator. However, the marginally reparameterized data in Fig.~\ref{fig:birds} show no residual bias and a stable estimation for many values of $k$ and $N/n$.  The figure can be generated by
\verb!NFkappaBDataBirdSpikingDataExample.m!, included in the distribution.

\section{\label{sec:discuss}Discussion}
While mutual information is being used routinely in analysis of modern
experimental data sets, high quality, unbiased estimation remains
an open problem. In this article, we described our modifications to the
well-known Kraskov, St{\"o}gbauer, and Grassberger
\cite{kraskov2004estimating} $k$ nearest neighbors estimator of mutual
information for real-valued data. Our contributions include developing
a method for estimating the variance of the estimator, for detecting
the presence of bias, and for choosing the optimal value of
$k$. Further, we suggest that transforming each marginal data dimension into the
standard normal form improves the range of applicability of the
estimator, allowing its use even for high-dimensional data sets. We substantiate our choices with extensive numerical
investigations. Finally, we provide a MatLab package implementing
these modifications to the KSG estimator, as well as a few examples
and a practical guide for the workflow. We hope that these
developments will be of use to a broad community of physics,
quantitative biology, and complex systems researchers.

We end this article with the following observation. As we mentioned in the {\em Introduction}, there are provably no universally unbiased estimators of mutual information, and thus every estimator---including the one we have developed here---will fail for some data sets. Nothing replaces looking at the data critically and thinking about whether the estimated values make sense and whether there are some patterns in the data that can be used to reduce the dimensionality, to simplify the estimation problem, or to verify the results. Blind application of any algorithm for estimation of mutual information in real-valued data, including application of our modification of the KSG approach, is likely to lead to a failure precisely when the data become interesting. 

\begin{acknowledgements}
We are thankful to Rachel Conn, Sam Sober, and other users of preliminary versions of our software packages for valuable feedback. We thank Raymond Cheong, Kyle Srivastava, Andre Levchenko, and Samuel Sober for providing experimental data for the examples in this work. CMH was supported in part by the Woodruff Scholarship at Emory University and the NSF Center for the Physics of Biological Function (PHY-1734030). IN was supported in part by NIH Grant 1R01NS099375 and NSF Grant IOS-1822677.
\end{acknowledgements}

\section*{\label{sec:appendix} Appendix}
We are trying to fit a model for the dependence of the KSG estimator variance on the sample size of the form 
\begin{equation}
\sigma^2_{\rm KSG}(N) = \left<\sigma^2_{\rm KSG}(N)\right>+{\rm noise}= \frac{B}{N} +{\rm noise},
\label{eq:var_form}
\end{equation}
where the angular brackets denote the expectation value.
By subsampling or partitioning the data, we can get (noisy) samples of the variance at smaller values $N_i$ than the actual maximum data set size, which we denote $N$. For each of these samples $\sigma^2_{\rm KSG}(N_i)\equiv\sigma^2_{{\rm KSG},i}$, $N_i = N/n_i$,  can be evaluated empirically, with  $n_i$ being the number of partitions of the data. For example, if we split the data into $n_i=3$ parts, we calculate the KSG mutual information for these 3 subsets, and we then estimate the variance at this $N_i$,  $\sigma^2_
{{\rm KSG}}(N_i)$ as the empirical variance of the three estimated values. Note that there can be multiple equal values of $n_i$ since data can be partitioned into the same number of parts in many different ways. 

The variable  $(n_i-1) \sigma^2_
{{\rm KSG},i}/\left<\sigma^2_{\rm KSG}(N_i)\right>$ obeys the $\chi^2$ distribution with $n_i-1$ degrees of freedom, $P^{(\chi^2)}_{n_i-1}(x)=\frac{1}{2^{(n_i-1)/2}\Gamma \left(\frac{n_i-1}{2}\right)}x^{\frac{n_i-1}{2}-1} e^{-x}$. Assuming independence of all $\sigma^2_{\rm KSG}(N_i)$ at different values of $i$, and using Eq.~(\ref{eq:var_form}), we view the product $\prod_i P^{(\chi^2)}_{n_i-1}\left(\frac{N(n_i-1) \sigma^2_{{\rm KSG},i}}{Bn_i}\right)$ as a likelihood function for $B$. Differentiating w.~r.~t.~$B$, we find the maximum likelihood (ML) solution
\begin{equation}
    B_{\rm ML}=\frac{\sum_i\frac{n_i-1}{n_i}N\sigma^2_{{\rm KSG}}(N_i)}{\sum_i(n_i-1)}.
\end{equation}
Thus the estimate of the KSG variance at the full data set size $N$ is
\begin{equation}
    \sigma^2_{\rm KSG}(N)=\frac{B}{N}=\frac{\sum_i\frac{n_i-1}{n_i}N\sigma^2_{{\rm KSG}}(N_i)}{\sum_i(n_i-1)}.
    \label{eq:var_est}
\end{equation}
We then calculate the standard error of $B$ and, with that, of the variance itself as the inverse of the second derivative of the log-likelihood at the maximum likelihood value:
\begin{equation}
    {\rm var}\, \sigma^2_{\rm KSG}(N)= \frac{2 B^2_{\rm ML}}{\sum_i(n_i-1)N^2}.
    \label{eq:var_var}
\end{equation}
These results are used for estimation of the KSG variance and its error bars in the main text. 

\bibliography{main}

\begin{thebibliography}{23}%
\makeatletter
\providecommand \@ifxundefined [1]{%
 \@ifx{#1\undefined}
}%
\providecommand \@ifnum [1]{%
 \ifnum #1\expandafter \@firstoftwo
 \else \expandafter \@secondoftwo
 \fi
}%
\providecommand \@ifx [1]{%
 \ifx #1\expandafter \@firstoftwo
 \else \expandafter \@secondoftwo
 \fi
}%
\providecommand \natexlab [1]{#1}%
\providecommand \enquote  [1]{``#1''}%
\providecommand \bibnamefont  [1]{#1}%
\providecommand \bibfnamefont [1]{#1}%
\providecommand \citenamefont [1]{#1}%
\providecommand \href@noop [0]{\@secondoftwo}%
\providecommand \href [0]{\begingroup \@sanitize@url \@href}%
\providecommand \@href[1]{\@@startlink{#1}\@@href}%
\providecommand \@@href[1]{\endgroup#1\@@endlink}%
\providecommand \@sanitize@url [0]{\catcode `\\12\catcode `\$12\catcode
  `\&12\catcode `\#12\catcode `\^12\catcode `\_12\catcode `\%12\relax}%
\providecommand \@@startlink[1]{}%
\providecommand \@@endlink[0]{}%
\providecommand \url  [0]{\begingroup\@sanitize@url \@url }%
\providecommand \@url [1]{\endgroup\@href {#1}{\urlprefix }}%
\providecommand \urlprefix  [0]{URL }%
\providecommand \Eprint [0]{\href }%
\providecommand \doibase [0]{http://dx.doi.org/}%
\providecommand \selectlanguage [0]{\@gobble}%
\providecommand \bibinfo  [0]{\@secondoftwo}%
\providecommand \bibfield  [0]{\@secondoftwo}%
\providecommand \translation [1]{[#1]}%
\providecommand \BibitemOpen [0]{}%
\providecommand \bibitemStop [0]{}%
\providecommand \bibitemNoStop [0]{.\EOS\space}%
\providecommand \EOS [0]{\spacefactor3000\relax}%
\providecommand \BibitemShut  [1]{\csname bibitem#1\endcsname}%
\let\auto@bib@innerbib\@empty
\bibitem [{\citenamefont {Shannon}\ and\ \citenamefont
  {Weaver}(1998)}]{shannon}%
  \BibitemOpen
  \bibfield  {author} {\bibinfo {author} {\bibfnamefont {C.}~\bibnamefont
  {Shannon}}\ and\ \bibinfo {author} {\bibfnamefont {W.}~\bibnamefont
  {Weaver}},\ }\href@noop {} {\emph {\bibinfo {title} {The mathematical theory
  of communication}}}\ (\bibinfo  {publisher} {University of Illinois Press},\
  \bibinfo {address} {Urbana, IL},\ \bibinfo {year} {1998})\BibitemShut
  {NoStop}%
\bibitem [{\citenamefont {Cover}\ and\ \citenamefont
  {Thomas}(2012)}]{cover2012elements}%
  \BibitemOpen
  \bibfield  {author} {\bibinfo {author} {\bibfnamefont {T.}~\bibnamefont
  {Cover}}\ and\ \bibinfo {author} {\bibfnamefont {J.}~\bibnamefont {Thomas}},\
  }\href@noop {} {\emph {\bibinfo {title} {Elements of information theory}}}\
  (\bibinfo  {publisher} {John Wiley \& Sons},\ \bibinfo {year}
  {2012})\BibitemShut {NoStop}%
\bibitem [{\citenamefont {Fairhall}\ \emph {et~al.}(2012)\citenamefont
  {Fairhall}, \citenamefont {Shea-Brown},\ and\ \citenamefont
  {Barreiro}}]{Fairhall:2012hk}%
  \BibitemOpen
  \bibfield  {author} {\bibinfo {author} {\bibfnamefont {A.}~\bibnamefont
  {Fairhall}}, \bibinfo {author} {\bibfnamefont {E.}~\bibnamefont
  {Shea-Brown}}, \ and\ \bibinfo {author} {\bibfnamefont {A.}~\bibnamefont
  {Barreiro}},\ }\href@noop {} {\bibfield  {journal} {\bibinfo  {journal} {Curr
  Opin Neurobiol}\ }\textbf {\bibinfo {volume} {22}},\ \bibinfo {pages} {653}
  (\bibinfo {year} {2012})}\BibitemShut {NoStop}%
\bibitem [{\citenamefont {Levchenko}\ and\ \citenamefont
  {Nemenman}(2014)}]{Levchenko:2014dy}%
  \BibitemOpen
  \bibfield  {author} {\bibinfo {author} {\bibfnamefont {A.}~\bibnamefont
  {Levchenko}}\ and\ \bibinfo {author} {\bibfnamefont {I.}~\bibnamefont
  {Nemenman}},\ }\href@noop {} {\bibfield  {journal} {\bibinfo  {journal} {Curr
  Opin Biotechn}\ }\textbf {\bibinfo {volume} {28C}},\ \bibinfo {pages} {156}
  (\bibinfo {year} {2014})}\BibitemShut {NoStop}%
\bibitem [{\citenamefont {Tkacik}\ and\ \citenamefont
  {Bialek}(2016)}]{Tkacik:2016ch}%
  \BibitemOpen
  \bibfield  {author} {\bibinfo {author} {\bibfnamefont {G.}~\bibnamefont
  {Tkacik}}\ and\ \bibinfo {author} {\bibfnamefont {W.}~\bibnamefont
  {Bialek}},\ }\href@noop {} {\bibfield  {journal} {\bibinfo  {journal} {Ann
  Rev Cond Matt Phys}\ }\textbf {\bibinfo {volume} {7}},\ \bibinfo {pages} {89}
  (\bibinfo {year} {2016})}\BibitemShut {NoStop}%
\bibitem [{\citenamefont {Paninski}(2003)}]{Paninski:2003vz}%
  \BibitemOpen
  \bibfield  {author} {\bibinfo {author} {\bibfnamefont {L.}~\bibnamefont
  {Paninski}},\ }\href@noop {} {\bibfield  {journal} {\bibinfo  {journal}
  {Neural Comput}\ }\textbf {\bibinfo {volume} {15}},\ \bibinfo {pages} {1191}
  (\bibinfo {year} {2003})}\BibitemShut {NoStop}%
\bibitem [{\citenamefont {Panzeri}\ and\ \citenamefont
  {Treves}(1996)}]{Panzeri:1996kv}%
  \BibitemOpen
  \bibfield  {author} {\bibinfo {author} {\bibfnamefont {S.}~\bibnamefont
  {Panzeri}}\ and\ \bibinfo {author} {\bibfnamefont {A.}~\bibnamefont
  {Treves}},\ }\href@noop {} {\bibfield  {journal} {\bibinfo  {journal}
  {Network}\ }\textbf {\bibinfo {volume} {7}},\ \bibinfo {pages} {87} (\bibinfo
  {year} {1996})}\BibitemShut {NoStop}%
\bibitem [{\citenamefont {Strong}\ \emph {et~al.}(1998)\citenamefont {Strong},
  \citenamefont {Koberle}, \citenamefont {de~Ruyter~van Steveninck},\ and\
  \citenamefont {Bialek}}]{strong1998entropy}%
  \BibitemOpen
  \bibfield  {author} {\bibinfo {author} {\bibfnamefont {S.}~\bibnamefont
  {Strong}}, \bibinfo {author} {\bibfnamefont {R.}~\bibnamefont {Koberle}},
  \bibinfo {author} {\bibfnamefont {R.}~\bibnamefont {de~Ruyter~van
  Steveninck}}, \ and\ \bibinfo {author} {\bibfnamefont {W.}~\bibnamefont
  {Bialek}},\ }\href@noop {} {\bibfield  {journal} {\bibinfo  {journal} {Phys
  Rev Lett}\ }\textbf {\bibinfo {volume} {80}},\ \bibinfo {pages} {197}
  (\bibinfo {year} {1998})}\BibitemShut {NoStop}%
\bibitem [{\citenamefont {Nemenman}\ \emph {et~al.}(2002)\citenamefont
  {Nemenman}, \citenamefont {Shafee},\ and\ \citenamefont
  {Bialek}}]{Nemenman:2002tm}%
  \BibitemOpen
  \bibfield  {author} {\bibinfo {author} {\bibfnamefont {I.}~\bibnamefont
  {Nemenman}}, \bibinfo {author} {\bibfnamefont {F.}~\bibnamefont {Shafee}}, \
  and\ \bibinfo {author} {\bibfnamefont {W.}~\bibnamefont {Bialek}},\ }in\
  \href@noop {} {\emph {\bibinfo {booktitle} {Adv Neural Inf Proc Syst
  (NIPS)}}},\ Vol.~\bibinfo {volume} {14},\ \bibinfo {editor} {edited by\
  \bibinfo {editor} {\bibfnamefont {T.}~\bibnamefont {Dietterich}}, \bibinfo
  {editor} {\bibfnamefont {S.}~\bibnamefont {Becker}}, \ and\ \bibinfo {editor}
  {\bibfnamefont {Z.}~\bibnamefont {Gharamani}}}\ (\bibinfo {year}
  {2002})\BibitemShut {NoStop}%
\bibitem [{\citenamefont {Panzeri}\ \emph {et~al.}(2007)\citenamefont
  {Panzeri}, \citenamefont {Senatore}, \citenamefont {Montemurro},\ and\
  \citenamefont {Petersen}}]{Panzeri:2007du}%
  \BibitemOpen
  \bibfield  {author} {\bibinfo {author} {\bibfnamefont {S.}~\bibnamefont
  {Panzeri}}, \bibinfo {author} {\bibfnamefont {R.}~\bibnamefont {Senatore}},
  \bibinfo {author} {\bibfnamefont {M.}~\bibnamefont {Montemurro}}, \ and\
  \bibinfo {author} {\bibfnamefont {R.}~\bibnamefont {Petersen}},\ }\href@noop
  {} {\bibfield  {journal} {\bibinfo  {journal} {J Neurophysiol}\ }\textbf
  {\bibinfo {volume} {98}},\ \bibinfo {pages} {1064} (\bibinfo {year}
  {2007})}\BibitemShut {NoStop}%
\bibitem [{\citenamefont {Zhang}(2012)}]{Zhang:2012ba}%
  \BibitemOpen
  \bibfield  {author} {\bibinfo {author} {\bibfnamefont {Z.}~\bibnamefont
  {Zhang}},\ }\href@noop {} {\bibfield  {journal} {\bibinfo  {journal} {Neural
  Comput}\ }\textbf {\bibinfo {volume} {24}},\ \bibinfo {pages} {1368–1389}
  (\bibinfo {year} {2012})}\BibitemShut {NoStop}%
\bibitem [{\citenamefont {Berry}\ \emph {et~al.}(2013)\citenamefont {Berry},
  \citenamefont {Tkacik}, \citenamefont {Dubuis}, \citenamefont {Marre},\ and\
  \citenamefont {da~Silveira}}]{Berry:2013tn}%
  \BibitemOpen
  \bibfield  {author} {\bibinfo {author} {\bibfnamefont {M.}~\bibnamefont
  {Berry}}, \bibinfo {author} {\bibfnamefont {G.}~\bibnamefont {Tkacik}},
  \bibinfo {author} {\bibfnamefont {J.}~\bibnamefont {Dubuis}}, \bibinfo
  {author} {\bibfnamefont {O.}~\bibnamefont {Marre}}, \ and\ \bibinfo {author}
  {\bibfnamefont {R.}~\bibnamefont {da~Silveira}},\ }\href@noop {} {\bibfield
  {journal} {\bibinfo  {journal} {J Stat Mech -Theory and Experiment}\ }\textbf
  {\bibinfo {volume} {2013}},\ \bibinfo {pages} {P03015} (\bibinfo {year}
  {2013})}\BibitemShut {NoStop}%
\bibitem [{\citenamefont {Archer}\ \emph {et~al.}(2014)\citenamefont {Archer},
  \citenamefont {Park},\ and\ \citenamefont {Pillow}}]{archer2014bayesian}%
  \BibitemOpen
  \bibfield  {author} {\bibinfo {author} {\bibfnamefont {E.}~\bibnamefont
  {Archer}}, \bibinfo {author} {\bibfnamefont {I.}~\bibnamefont {Park}}, \ and\
  \bibinfo {author} {\bibfnamefont {J.}~\bibnamefont {Pillow}},\ }\href@noop {}
  {\bibfield  {journal} {\bibinfo  {journal} {J Machine Learning Res}\ }\textbf
  {\bibinfo {volume} {15}},\ \bibinfo {pages} {2833} (\bibinfo {year}
  {2014})}\BibitemShut {NoStop}%
\bibitem [{\citenamefont {Ma}(1981)}]{Ma-1981}%
  \BibitemOpen
  \bibfield  {author} {\bibinfo {author} {\bibfnamefont {S.}~\bibnamefont
  {Ma}},\ }\href@noop {} {\bibfield  {journal} {\bibinfo  {journal} {J Stat
  Phys}\ }\textbf {\bibinfo {volume} {26}},\ \bibinfo {pages} {221} (\bibinfo
  {year} {1981})}\BibitemShut {NoStop}%
\bibitem [{\citenamefont {Tang}\ \emph {et~al.}(2014)\citenamefont {Tang},
  \citenamefont {Chehayeb}, \citenamefont {Srivastava}, \citenamefont
  {Nemenman},\ and\ \citenamefont {Sober}}]{tang2014millisecond}%
  \BibitemOpen
  \bibfield  {author} {\bibinfo {author} {\bibfnamefont {C.}~\bibnamefont
  {Tang}}, \bibinfo {author} {\bibfnamefont {D.}~\bibnamefont {Chehayeb}},
  \bibinfo {author} {\bibfnamefont {K.}~\bibnamefont {Srivastava}}, \bibinfo
  {author} {\bibfnamefont {I.}~\bibnamefont {Nemenman}}, \ and\ \bibinfo
  {author} {\bibfnamefont {S.}~\bibnamefont {Sober}},\ }\href@noop {}
  {\bibfield  {journal} {\bibinfo  {journal} {PLoS biology}\ }\textbf {\bibinfo
  {volume} {12}},\ \bibinfo {pages} {e1002018} (\bibinfo {year}
  {2014})}\BibitemShut {NoStop}%
\bibitem [{\citenamefont {Srivastava}\ \emph {et~al.}(2017)\citenamefont
  {Srivastava}, \citenamefont {Holmes}, \citenamefont {Vellema}, \citenamefont
  {Pack}, \citenamefont {Elemans}, \citenamefont {Nemenman},\ and\
  \citenamefont {Sober}}]{srivastava2017motor}%
  \BibitemOpen
  \bibfield  {author} {\bibinfo {author} {\bibfnamefont {K.}~\bibnamefont
  {Srivastava}}, \bibinfo {author} {\bibfnamefont {C.}~\bibnamefont {Holmes}},
  \bibinfo {author} {\bibfnamefont {M.}~\bibnamefont {Vellema}}, \bibinfo
  {author} {\bibfnamefont {A.}~\bibnamefont {Pack}}, \bibinfo {author}
  {\bibfnamefont {C.}~\bibnamefont {Elemans}}, \bibinfo {author} {\bibfnamefont
  {I.}~\bibnamefont {Nemenman}}, \ and\ \bibinfo {author} {\bibfnamefont
  {S.}~\bibnamefont {Sober}},\ }\href@noop {} {\bibfield  {journal} {\bibinfo
  {journal} {Proc Natl Acad Sci (USA)}\ }\textbf {\bibinfo {volume} {114}},\
  \bibinfo {pages} {1171} (\bibinfo {year} {2017})}\BibitemShut {NoStop}%
\bibitem [{\citenamefont {Kraskov}\ \emph {et~al.}(2004)\citenamefont
  {Kraskov}, \citenamefont {St{\"o}gbauer},\ and\ \citenamefont
  {Grassberger}}]{kraskov2004estimating}%
  \BibitemOpen
  \bibfield  {author} {\bibinfo {author} {\bibfnamefont {A.}~\bibnamefont
  {Kraskov}}, \bibinfo {author} {\bibfnamefont {H.}~\bibnamefont
  {St{\"o}gbauer}}, \ and\ \bibinfo {author} {\bibfnamefont {P.}~\bibnamefont
  {Grassberger}},\ }\href@noop {} {\bibfield  {journal} {\bibinfo  {journal}
  {Phys Rev E}\ }\textbf {\bibinfo {volume} {69}},\ \bibinfo {pages} {066138}
  (\bibinfo {year} {2004})}\BibitemShut {NoStop}%
\bibitem [{\citenamefont {Kozachenko}\ and\ \citenamefont
  {Leonenko}(1987)}]{kozachenko1987sample}%
  \BibitemOpen
  \bibfield  {author} {\bibinfo {author} {\bibfnamefont {L.}~\bibnamefont
  {Kozachenko}}\ and\ \bibinfo {author} {\bibfnamefont {N.~N.}\ \bibnamefont
  {Leonenko}},\ }\href@noop {} {\bibfield  {journal} {\bibinfo  {journal}
  {Problemy Peredachi Informatsii}\ }\textbf {\bibinfo {volume} {23}},\
  \bibinfo {pages} {9} (\bibinfo {year} {1987})}\BibitemShut {NoStop}%
\bibitem [{\citenamefont {St{\"o}gbauer}\ \emph {et~al.}(2004)\citenamefont
  {St{\"o}gbauer}, \citenamefont {Kraskov}, \citenamefont {Astakhov},\ and\
  \citenamefont {Grassberger}}]{stogbauer2004least}%
  \BibitemOpen
  \bibfield  {author} {\bibinfo {author} {\bibfnamefont {H.}~\bibnamefont
  {St{\"o}gbauer}}, \bibinfo {author} {\bibfnamefont {A.}~\bibnamefont
  {Kraskov}}, \bibinfo {author} {\bibfnamefont {S.}~\bibnamefont {Astakhov}}, \
  and\ \bibinfo {author} {\bibfnamefont {P.}~\bibnamefont {Grassberger}},\
  }\href@noop {} {\bibfield  {journal} {\bibinfo  {journal} {Phys Rev E}\
  }\textbf {\bibinfo {volume} {70}},\ \bibinfo {pages} {066123} (\bibinfo
  {year} {2004})}\BibitemShut {NoStop}%
\bibitem [{\citenamefont {Efron}\ and\ \citenamefont
  {Tibshirani}(1993)}]{Efron:1993tv}%
  \BibitemOpen
  \bibfield  {author} {\bibinfo {author} {\bibfnamefont {B.}~\bibnamefont
  {Efron}}\ and\ \bibinfo {author} {\bibfnamefont {R.}~\bibnamefont
  {Tibshirani}},\ }\href@noop {} {\emph {\bibinfo {title} {An introduction to
  the bootstrap}}}\ (\bibinfo  {publisher} {Chapman \& Hall},\ \bibinfo
  {address} {New York},\ \bibinfo {year} {1993})\BibitemShut {NoStop}%
\bibitem [{\citenamefont {Nemenman}\ \emph {et~al.}(2008)\citenamefont
  {Nemenman}, \citenamefont {Lewen}, \citenamefont {Bialek},\ and\
  \citenamefont {de~Ruyter~van Steveninck}}]{Nemenman:2008ft}%
  \BibitemOpen
  \bibfield  {author} {\bibinfo {author} {\bibfnamefont {I.}~\bibnamefont
  {Nemenman}}, \bibinfo {author} {\bibfnamefont {G.}~\bibnamefont {Lewen}},
  \bibinfo {author} {\bibfnamefont {W.}~\bibnamefont {Bialek}}, \ and\ \bibinfo
  {author} {\bibfnamefont {R.}~\bibnamefont {de~Ruyter~van Steveninck}},\
  }\href@noop {} {\bibfield  {journal} {\bibinfo  {journal} {PLoS Comput Biol}\
  }\textbf {\bibinfo {volume} {4}},\ \bibinfo {pages} {e1000025} (\bibinfo
  {year} {2008})}\BibitemShut {NoStop}%
\bibitem [{\citenamefont {Miller}(1955)}]{miller}%
  \BibitemOpen
  \bibfield  {author} {\bibinfo {author} {\bibfnamefont {G.}~\bibnamefont
  {Miller}},\ }in\ \href@noop {} {\emph {\bibinfo {booktitle} {Information
  Theory in Psychology II-B}}},\ \bibinfo {editor} {edited by\ \bibinfo
  {editor} {\bibfnamefont {H.}~\bibnamefont {Quastler}}}\ (\bibinfo
  {publisher} {Free Press},\ \bibinfo {address} {Glencoe, IL},\ \bibinfo {year}
  {1955})\ pp.\ \bibinfo {pages} {95--100}\BibitemShut {NoStop}%
\bibitem [{\citenamefont {Cheong}\ \emph {et~al.}(2011)\citenamefont {Cheong},
  \citenamefont {Rhee}, \citenamefont {Wang}, \citenamefont {Nemenman},\ and\
  \citenamefont {Levchenko}}]{Cheong:2011jp}%
  \BibitemOpen
  \bibfield  {author} {\bibinfo {author} {\bibfnamefont {R.}~\bibnamefont
  {Cheong}}, \bibinfo {author} {\bibfnamefont {A.}~\bibnamefont {Rhee}},
  \bibinfo {author} {\bibfnamefont {C.}~\bibnamefont {Wang}}, \bibinfo {author}
  {\bibfnamefont {I.}~\bibnamefont {Nemenman}}, \ and\ \bibinfo {author}
  {\bibfnamefont {A.}~\bibnamefont {Levchenko}},\ }\href@noop {} {\bibfield
  {journal} {\bibinfo  {journal} {Science}\ }\textbf {\bibinfo {volume}
  {334}},\ \bibinfo {pages} {354} (\bibinfo {year} {2011})}\BibitemShut
  {NoStop}%
\end{thebibliography}%
\end{document}